\documentclass{emulateapj}

\usepackage{natbib}

\newcommand{\kms}{km\,s$^{-1}$}
\newcommand{\cms}{cm\,s$^{-1}$}
\newcommand{\kmskpc}{km\,s$^{-1}$\,kpc$^{-1}$}
\newcommand{\Msun}{M$_\odot$}

\newcommand{\Msunyr}{M$_\odot$\,yr$^{-1}$}
\newcommand{\rs}{r_s} 
\newcommand{\Ms}{M_s} 
\newcommand{\rd}{r_d} 
\newcommand{\hd}{h_d} 
\newcommand{\rdz}{r_{d,0}} 
\newcommand{\rde}{r_{d,edge}} 
\newcommand{\Md}{M_d} 
\newcommand{\rHd}{r_{H,d}} 
\newcommand{\TTdf}{T_\mathrm{df}} 
\newcommand{\TTd}{T_\mathrm{disk}} 
\newcommand{\TTi}{T_\mathrm{infl}} 
\newcommand{\TTv}{T_\mathrm{visc}} 
\newcommand{\Menc}{M_\mathrm{enc}} 
\newcommand{\Mencs}{M_\mathrm{enc,s}} 
\newcommand{\Mencz}{M_\mathrm{enc,0}} 
\newcommand{\OFs}{\Omega_s} 
\newcommand{\tcr}{t_\mathrm{cross}} 
\newcommand{\tdf}{t_\mathrm{df}} 
\newcommand{\tdfz}{t_\mathrm{df,0}} 
\newcommand{\tsep}{t_\mathrm{sep}} 
\newcommand{\qd}{q_d} 
\newcommand{\qp}{q'} 
\newcommand{\alp}{\alpha'} 
\newcommand{\bp}{\beta'} 
\newcommand{\gp}{\gamma'} 
\newcommand{\Dr}{\Delta r} 
\newcommand{\vorb}{v_\mathrm{orb}} 
\newcommand{\Mc}{M_c} 
\newcommand{\rion}{r_\mathrm{ion}} 
\newcommand{\Mstar}{M_\star} 
\newcommand{\du}{\mathrm{d}} 
\newcommand{\HH}{\ensuremath{\mathrm{H}_2}}
\newcommand{\Ha}{\ensuremath{\mathrm{H}\alpha}}

\newcommand{\NII}{\textsc{N}\,\textsc{II}}

\shorttitle{Migration of Star Clusters and Nuclear Rings}
\shortauthors{van de Ven \& Chang}

\begin{document}

\title{Migration of Star Clusters and Nuclear Rings}

%
%
%
%
%
%
%
%

\author{Glenn van de Ven\altaffilmark{1,3}}
\author{Philip Chang\altaffilmark{2,4}}
\affil{$^1$Institute for Advanced Study, Einstein Drive, Princeton, NJ
  08540, USA; glenn@ias.edu} 
\affil{$^2$Astronomy Department and Theoretical Astrophysics Center, 601
  Campbell Hall, University of California, Berkeley, CA 94720; pchang@astro.berkeley.edu} 
\altaffiltext{3}{Hubble Fellow}
\altaffiltext{4}{Miller Institute for Basic Research}

\begin{abstract}
  Star clusters that form in nuclear rings appear to be at slightly
  larger radii than the gas. We argue that the star clusters move out
  from the gas in which they are formed because of satellite-disk
  tidal interactions. In calculating the dynamics of this star cluster
  and gas ring system, we include the effects of dynamical friction of
  the background stars in the host galaxy on the star cluster, and
  inflowing gas along the bar onto the nuclear ring at the two contact
  points. We show that the final separation is of the order of the
  Hill radius of the nuclear ring, which is typically $20-30\%$ of its
  radius. Massive star clusters can reach half of this separation very
  quickly and produce a factor of a few enhancement in the gas surface
  density. If this leads to star formation in addition to the
  (ongoing) formation of star clusters near the contact points, a
  possible (initial) azimuthal age gradient may become diluted or even
  disappear. Finally, if the star cluster are massive and/or numerous
  enough, we expect the nuclear ring to migrate inward, away from the
  (possibly) associated (inner) Lindblad resonance. We discuss how
  these predictions may be tested observationally.
\end{abstract}


\section{Introduction}
\label{sec:intro}

Nuclear rings in barred spiral galaxies are regions of large gas
surface densities and high star formation. Their existence is
intimately related to the stellar bars with which they are associated.
The stellar bar drives the interstellar medium (ISM) into the nuclear
ring.  How the radius of the ring is determined is unclear. The radius
may be set by the location of the inner Lindblad resonance (ILR) of
the stellar bar \citep[see review by][and references
therein]{Buta1996}. The reasoning behind this idea is that the sign of
the torque changes on either side of the Lindblad resonance.
Alternatively, the radius may be at the location where the
non-axisymmetric bar weakens, which is typically at $\sim 10\%$ of the
size of the bar \citep{Shlosman1990}. Finally, it may also be
determined by where the transition from phase space domination of
$X_1$ orbits (parallel to the major axis of the large-scale bar) to
the phase space domination of $X_2$ orbits (perpendicular to the bar)
occurs \citep{Regan1997}.

Regardless of how they are formed they are interesting as a class
because they are perhaps the largest population of nearby star
bursting regions.  They form star clusters in prodigious amounts and
are the only environment where super star clusters (SSCs) might be
found in abundance in normal galaxies \citep{Maoz2001}.  These SSCs
can have masses in excess of $10^6$\,\Msun\ and may be the progenitors
of globular clusters if they can survive for $10^{10}$ years.

In some of these systems the location of these star clusters is
curious. They appear to be at larger radii than the gas ring from
which they are presumably formed. For instance, the morphology of
NGC\,1512 clearly show the star cluster(s) exterior (radially) to the gas
as opposed to being embedded in the gas \citep{Maoz2001}. Taking
another example, in NGC\,4314 the star clusters are at larger radii
than the gas in the nuclear ring \citep{Benedict2002}. Finally,
\cite{Martini2003a} find that in their sample of 123 galaxies, eight
have strong nuclear rings and in each of the eight, star formation
occurs outside of the dust ring. Two of their other galaxies have
dust lanes immediately interior to what appears to be an older
stellar ring.

There are two possible explanations for why these star clusters are
external to the gas. First, while the gas can lose angular momentum
due to dissipation, the stars and star clusters are collisionless and
hence remain on larger orbits \citep{Regan2003}. Second, the star
clusters migrated outward after their formation. We explore the latter
possibility in this paper.

We argue that tidal interactions between the star cluster and the
nuclear ring from which it was formed leads to their separation. We
apply the physics of ``gas shepherding'', which was studied by one of
us \citep[][hereafter C08]{Chang2008}, to this cluster-ring system. We
derive a natural length scale for this separation, which we call the
``Hill radius'' of the ring. At the same time, we show that the outer
edge of the ring can experience a surface density enhancement of a
few, possibly leading to additional formation of stars and star
clusters along the edge of the ring.  Finally, we find that if the
star clusters are massive and/or numerous enough the cluster-ring
systems as a whole is expected to migrate inward. We show that these
predictions can be tested observationally.

We present our model in \S~\ref{sec:tidal}. After summarizing the
observed properties of nuclear rings in \S~\ref{sec:properties}, we
give the basic picture and introduce the Hill radius of the ring in
\S~\ref{sec:basic_picture}, followed by the basic equations of our
model in \S~\ref{sec:basic_equations}. The solutions for various
scenarios are outlined in \S~\ref{sec:results}. A few observationally
testable predictions that arises from our model are given and compared
to observations in \S~\ref{sec:comparison}. We discuss some aspects of
our model in \S~\ref{sec:discussion} and summarize our conclusions in
\S~\ref{sec:conclusions}.

\section{Tidal Interactions Between Star Clusters and Nuclear Rings}
\label{sec:tidal}

We first summarize the basic properties of observed nuclear rings,
which are used to place the model predictions in context. We then
present the basic picture of our model, followed by the basic set of
equations that govern our model.

\subsection{Observed properties}
\label{sec:properties}

For a sample of 22 nuclear rings, \cite{Mazzuca2008} find that the
rings are in the same plane as the disk of their host galaxy and that
they are intrinsically circular. Typically, the radius of the ring is
around $\sim 0.5$\,kpc, but it can be a factor three smaller or
larger. With a circular velocity at the ring radius of around $\sim
150$\,\kms, the (spherically) enclosed mass is about $\Menc \sim 2.5
\times 10^9$\,\Msun.

The gaseous mass in the ring may be estimated from high spatial
resolution CO measurements, using the 'standard' conversion to \HH\ 
mass and adding $36\%$ for He. In this way, we obtain for NGC\,5457,
NGC\,3351, NGC\,6951, NGC\,3504 \citep{Kenney1992,Kenney1993} and
NGC\,4314 \citep{Benedict1996} on average a mass inside the ring of
$\Md \sim 5 \times 10^7$\,\Msun.  Although the ring mass may be as
much as a factor ten smaller or two larger, it correlates with the
ring radius, so that the ring-to-enclosed mass ratio $\qd \equiv
\Md/\Menc$ stays within a narrow range of $\qd = 0.01 - 0.03$.
While this is the molecular gas inside the ring, part of the gas is
being ionized, due to photons from newly formed massive stars
or due to shocks from stellar winds or supernovae explosions, and at
the inner edge of the ring possibly even due to an active galactic
nucleus (AGN). However, the ionized gas mass estimated from the \Ha\ 
(plus [\NII]) emission is generally two orders of magnitude lower than
the molecular gas mass \citep[e.g.][]{Planesas1997}.

The nuclear rings are easily observable due to the emission from the
massive stars inside the young star clusters in the ring (which in the
ultraviolet even dominates the emission from a possible AGN). The
combination of high-spatial resolution (HST) images and
multi-wavelength coverage, makes it possible to isolate the individual
clusters and after correction for (dust) extinction to estimate their
mass and (relative) age. Below we use such a study by
\cite{Benedict2002} to investigate in more detail the nuclear ring in
NGC\,4314. Overall, the cluster masses follow a power-law distribution
with index $-2$, similar to young star clusters in merging galaxies
such as the Antennae \citep{Zhang1999}, but typically extending to a
smaller upper limit in mass (few times $10^5$\,\Msun), i.e., fewer of
the so-called ``super star clusters'' \citep[but see][]{Maoz2001}.

On the other hand, these young star clusters are most likely just the
latest star formation in an ongoing succession of bursts
\citep[e.g.][]{Allard2006}. Most of the lower-mass clusters dissolve,
either early-on ($\la 50$\,Myr) due to violent relaxation after
expelling the residual gas ('infant mortality') depending on the star
formation efficiency \citep[e.g.][]{Parmentier2008}, or later-on due
to internal relaxation ('evaporation') and external tidal effects
\citep[e.g.][]{McLaughlin2007}. The resulting evolution towards a
bell-shaped distribution implies a relative increase in more massive
star clusters over time. Unfortunately, once the most massive stars
have died, the luminosity of star clusters very rapidly fades, so that
it becomes more difficult to distinguish them from the old stellar
population of their host galaxy. Nevertheless, based on velocity
dispersion measurements from stellar absorption lines,
\cite{Hagele2007} estimate dynamical masses up to almost $\sim
10^7$\,\Msun\ for individual star clusters in the nuclear ring of
NGC\,3351. Henceforth, the typical ratio of cluster-to-ring mass
ratios $\qp \equiv \Ms/\Md$ depends on the (average) age of the star
clusters. For the young, luminous star clusters we adopt the range
$\qp = 0.001 - 0.01$, whereas for the old(er), faint(er) star cluster
$\qp = 0.01 - 0.1$ is probably more appropriate. Below we investigate
the two cases $\qp = 0.1$ and $\qp = 0.01$, corresponding to $\Ms \sim
5 \times 10^6$\,\Msun\ and $\Ms \sim 5 \times 10^5$\,\Msun\ in case of
the above average gas mass in nuclear rings of $\Md \sim 5 \times
10^7$\,\Msun. The effect of a less massive star cluster with $\qp <
0.01$ is similar to that of $\qp = 0.01$.

The time for a star cluster to cross the nuclear ring is given by
$\tcr = 2\pi/\Omega$, where $\Omega = \sqrt{G\Menc/r^3}$ is the
orbital frequency and $G$ is Newton's constant of gravity. Given a
typical circular velocity of $\sim 150$\,\kms\ at the observed mean
nuclear ring radius of $0.5$\,kpc, the crossing time is around $\tcr
\sim 20$\,Myr. A star cluster is subject to dynamical friction with
the bulge stars on a (minimum) timescale\footnote{The dynamical
  friction time is likely to be longer for a partially rotationally
  supported bulge as opposed to a non-rotating bulge.} $\tdf = \Menc /
(\Omega\Ms\ln\Lambda)$, with $\ln\Lambda$ the Coulomb logarithm. Using
$\tcr = \tdf \, \qp \qd \, 2\pi \ln\Lambda$ and substituting $\qd
\simeq 0.02$ from above, we find that the dynamical friction time in
units of the crossing time is approximately given by $\tdf/\tcr \simeq
1/\qp$. This means hundreds to tens of orbits for star clusters with
masses that are $1\%$ ($\qp=0.01$) to $10\%$ ($\qp=0.1$) of the
nuclear ring mass, or, with $\tcr = 20$\,Myr, a dynamical friction
time $\tdf$ from about $2$\,Gyr to $200$\,Myr for a non-rotating
bulge, which we will assume for the remainder of this work.

\subsection{Basic Picture}
\label{sec:basic_picture}

\begin{figure*}
  \plotone{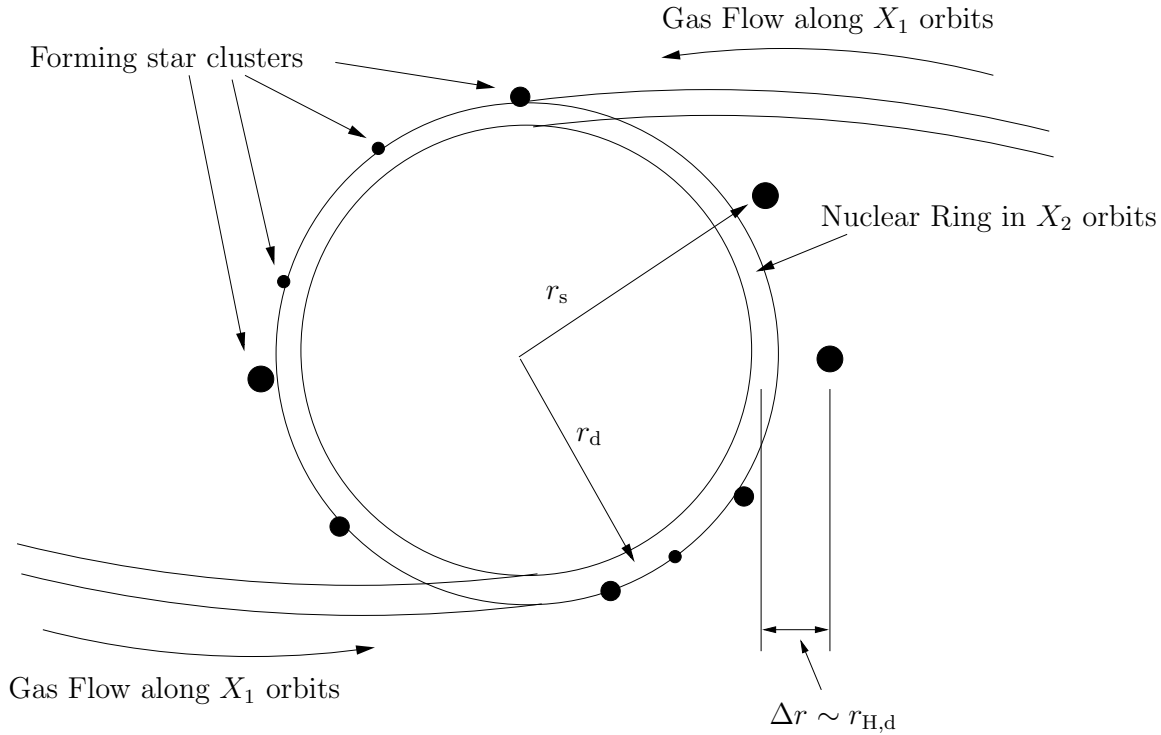}
\caption{Cartoon of our model. Gas flows from larger radii
  (along a $X_1$ orbit parallel to the major axis of the large-scale
  bar) onto the nuclear ring, which is in an approximately circular
  $X_2$ orbit. Part of this gas collapses to form star clusters near
  the outer edge of the ring. These star clusters, being subject to
  dynamical friction with the surrounding bulge stars in the host galaxy,
  tidally interact with the gas in the nuclear ring and migrate
  outward to a radius $\rs$. At the same time, the star clusters that
  are sufficiently massive push the outer edge of the nuclear ring to
  a smaller radius $\rd$. The net separation between the star clusters
  and nuclear ring, $\Dr = \rs - \rd$, is of order the ``Hill radius''
  of the nuclear ring, $\rHd$.}
\label{fig:cartoon}
\end{figure*}

We illustrate our basic model in a cartoon in Figure
\ref{fig:cartoon}. Gas flowing in a bar potential along the $X_1$
orbits transitions to $X_2$ orbits making up the nuclear ring.  The
$X_2$ orbits are approximately circular and so we will approximate
them as circles. From the gas in the nuclear ring stars are formed in
star clusters. We assume that the star clusters are formed
preferentially near the outer edge. These newly formed and forming
star clusters are subject to dynamical friction with the background
bulge stars\footnote{The background bulge stars may also be halo
  stars, and the bulge/halo is assumed not to be supported by
  rotation. For the purposes of this work, we will idealize the
  bulge/halo as an isothermal background distribution, which we argue
  in \S~\ref{sec:discussion_isothermal_profile} is a valid
  approximation.} and tidal interactions with the ring. However, the
initial tidal interactions are strong compared to dynamical friction
and thus these star clusters will initially move {\it away} from the
ring.

Indeed, in observations of nuclear rings young star clusters often
appear to be at larger radii than the gas ring from which they are
presumably formed. The color composite images of NGC\,1512
\citep{Maoz2001}, NGC\,4314 \citep{Benedict2002}, and NGC\,7742
(Hubble Heritage\footnote{http://heritage.stsci.edu/1998/28/}), show
young star cluster as bright spots on the outside of gas traced by the
dust.  As pointed out by the referee, from these images it is also
clear that, even though the nuclear rings are as we assume
approximately circular with star clusters at the outer edge, the
observations can be quite difficult and complex.
First, when resolved in the optical the rings can look like tightly
wound spirals. This might be just appearance because the bar dust
lanes connecting with the ring are also partly obscuring the ring in
optical broad-band images, whereas in e.g.\ narrow-band imaging around
\Ha\ the ring of young star clusters seem to be complete \citep[see
also][]{Maoz2001}. Nevertheless, our basic picture described above for
a circular ring should still provide a fair description in case of a
tightly wound spiral.
Second, very obscured star clusters could remain hidden below the gas
ring even in the near infrared images taken, whereas even longer
wavelength observations (yet) lack the high spatial resolution
required to resolve the star clusters. Even so, the clearing of gas
and dust from the newly formed star cluster is very fast and
efficient, as even the youngest visible star clusters are only mildly
reddened \citep[e.g.][]{Maoz2001}.
Finally, some (sectors of) star clusters seem to appear more on the
inside of the nuclear ring, possibly along an inward extension of the
gas inspiraling onto the ring along the large-scale bar. As we
discuss briefly in \S~\ref{sec:discussion_star_formation}, star
clusters may form within the ring, in particular at the contact points
where the gas flows into the ring. Moreover, the interaction of
non-axisymmetric structures inside the ring (such as nuclear bars and
spirals) with the interstellar medium may also induce the formation of
star clusters.
However, even though such additional mechanism might be operating and
the observations should be interpret with care, they in general
support our basic picture of a circular nuclear ring with young stars
migrating outward.

Since the basic physics mirror the "gas shepherding" scenario studied
by C08, we adopt the same notation. In this case the satellite is a
star cluster and the disk is a ring, but since only the (local)
geometry at the edge matters the physics remains the same.  There are
however two important differences with respect to C08.  First, the
perturbing star cluster's initial radial position begins very close to
the ring as opposed to a satellite reaching the disk from very far
away in C08.  Due to ring tidal interactions, the star cluster and
ring will separate to establish an initial separation, $\Delta r$,
which we show below is of order the ``Hill radius'' of the ring. We
aim to study the manner by which this initial separation is achieved.
Second, gas continues to flow in from the $X_1$ orbits so that at the
transition between the $X_1$ to $X_2$ orbits, gas is continuously
being added to the system. Therefore, a source of mass exists in this
system, which changes the dynamics.

\placefigure{fig:cartoon}

For now we ignore the source of mass to concentrate just on the tidal
interaction between the ring and star cluster and the dynamical
friction experienced by the star cluster. We begin by showing that
this combined action of dynamical friction and ring tides results in a
length scale in the problem, which we call the Hill radius of the
ring.
The torque due to dynamical friction on the star cluster is
\citep{Chandrasekhar1943, BT87}
\begin{equation}\label{eq:torque_estimate_df}
  \TTdf \sim \frac{\Ms}{\tdf} \, \OFs \rs^2
  \sim \frac{G \Ms^2}{\rs},
\end{equation}
where $\Ms$ is the mass of the star cluster, $\rs$ is its distance
from the center of the galaxy, $\OFs$ is its orbital frequency, and,
as before, $\tdf$ is the timescale for dynamical friction.
The torque due to the excitation of spiral density waves in a disk is
\citep{Goldreich1980, Lin1986, Ward1989, Artymowicz1993, Ward1997}
\begin{equation}\label{eq:torque_estimate_disk}
  \TTd \sim \frac{G\Ms^2}{\rs} \frac{\Md}{\Mencs} 
  \left(\frac{\rs}{\Dr}\right)^3,   
\end{equation}
where $\Md$ is the mass of the ring with radius $r_d$. The radial
separation between the star cluster and the ring $\Dr = \rs - \rd$ is
assumed to be small compared to the distance to the center, i.e., $\Dr
\ll \rd < \rs$.
While dynamical friction torques down the star cluster, the ring
torques up the star cluster. Balancing these two torques, i.e., $\TTdf
\sim \TTd$, gives
\begin{equation}\label{eq:hill_radius}
  \Dr \sim \rs \left(\frac{\Md}{\Mencs}\right)^{1/3} \equiv \rHd,
\end{equation}
where the latter defines the Hill radius of the ring. 

Given the observed range in ring-to-enclosed mass of $\qd = 0.01 -
0.03$ (\S~\ref{sec:properties}), the Hill radius is around $20 - 30\%$
of the radius $\rs$ of the star cluster. Since $\rs \sim \rd \gg
\rHd$, the observed range in nuclear ring radii of $\rd = 0.2 -
1.7$\,kpc \citep[e.g.][]{Mazzuca2008} results in $\rHd = 50 -
500$\,pc. We expect the radial separation between the star cluster and
nuclear ring to be of order of this Hill radius. However, the exact
amount and nature by which this separation will be achieved depends
strongly on the strength of the tidal torque between the cluster and
the ring, which we intend to quantify in this paper.

The second process which we wish to study is the effect of an
additional source of mass on the dynamics of the cluster-ring system.
The gas merges onto the nuclear ring at the two contact points where
the $X_1$ orbits transition to $X_2$ orbits. We assume that the region
where the gas is added to the nuclear ring is radially narrow compared
to the radius of the nuclear ring. For a mass inflow rate of
$\dot{M}$, the amount of angular momentum added to a system per unit
time is
\begin{equation}\label{eq:dotL_inflow}
  \dot{L} = \dot{M} \, \Omega r^2.
\end{equation}
However, not all of this angular momentum can be immediately applied
to affecting the qualitative dynamics of the system. By qualitative,
we mean that the mass inflow rate, which acts as a source of angular
momentum, changes the behavior of the cluster-ring system. For
instance, the source of angular momentum from mass inflow balances the
sink of angular momentum from dynamical friction on the star cluster.
To estimate what mass inflow rate would be needed to effect such a
change, we suppose this mass is added just inside of the satellite
radial position, i.e., $r \simeq \rs$.  The mass will then be pushed
to the edge of the ring, $\rd$, due to tidal interactions. The change
in radial position will be of order the Hill radius of the disk,
$\rHd$. Hence, for such a parcel of gas of mass $M$, the change in
angular momentum moving from $\rs$ to $\rd$ is
\begin{equation}\label{eq:deltaL_estimate_inflow}
  \Delta L \sim M \, \Omega r \, \Dr \sim M \, \OFs \rs \, \rHd.
\end{equation}
For a constant mass inflow rate $\dot{M}$, we expect the torque from
this inflow of gas, acting on the star cluster, to be
\begin{equation}\label{eq:torque_estimate_inflow}
  \TTi = \frac{\du \Delta L}{\du t}
  \sim \dot{M} \, \OFs \rs \, \rHd.
\end{equation}
Setting this inflow torque equal to the dynamical friction torque in
eq.~(\ref{eq:torque_estimate_df}), we find the scale for the mass
inflow rate to be
\begin{equation}\label{eq:mass_inflow_scale}
  \dot{M} \sim \frac{\Ms}{\tdf} \left(\frac{\Mencs}{\Md}\right)^{1/3},
\end{equation}
where we have used the definition of the Hill radius of the ring given
in eq.~(\ref{eq:hill_radius}). Had we instead (naively) set
eq.~(\ref{eq:dotL_inflow}) to be equal to
eq.~(\ref{eq:torque_estimate_df}), we would have found $\dot{M} \sim
\Ms/\tdf$, which can be significantly smaller than the more physical
scale in eq.~(\ref{eq:mass_inflow_scale}).

As shown in \S~\ref{sec:properties} above, $\tdf \simeq \tcr/\qp$, so
that $\dot{M} \sim (\Md/\tcr) \, \qp^2 / \qd^{1/3}$. Substituting the
average (gas) mass inside the ring of $\Md \sim 5 \times 10^7$\,\Msun,
the crossing time of $\tcr \sim 20$\,Myr, and the observed range in
ring-to-enclosed mass of $\qd = 0.01 - 0.03$, the typical scale for
the mass inflow rate becomes $\dot{M} \sim 10 \, \qp^2$\,\Msunyr.  For
star clusters with masses that are $1\%$ ($\qp=0.01$) to $10\%$
($\qp=0.1$) of the nuclear ring mass, this implies a scale $\dot{M}$
from about $0.001$ to $0.1$\,\Msunyr. To balance the \emph{combined}
dynamical friction from the many star clusters that are formed, the
\emph{total} mass inflow rate might easily have to be as high as $\sim
1$\,\Msunyr, which is the typical star formation rate in nuclear rings
\citep[e.g.][]{Mazzuca2008}.  For mass inflow rates significantly
below this scale, we expect the effect of this source of (gas) mass to
be insignificant. For mass inflow rates at or above this scale, the
effect of dynamical friction to push the cluster-ring system as a
whole to smaller radii is \emph{halted}.

\subsection{Basic Equations}
\label{sec:basic_equations}

We now present the governing equations for our model.  We do not
present a detailed and complete derivation because these equations
have been already derived in much of the literature
\citep[e.g.][]{Lin1979a, Lin1979b, Lin1986, Hourigan1984, Ward1989,
  Rafikov2002}, with two differences. First, we include the effect of
dynamical friction of the background bulge stars, which is extensively
discussed in C08 (to which we refer the interested reader). Second, we
will include a mass source term for, which models the inflow of gas
mass transitioning from the $X_1$ to $X_2$ orbits.

The evolution of a \emph{viscous} gas ring under the influence of an
external torque is given by equations for continuity and angular
momentum. The equation of continuity is
\begin{equation}\label{eq:continuity}
  \frac{\partial \Sigma}{\partial t} 
  + \frac{1}{r} \frac{\partial (r v_r \Sigma)}{\partial r} = S(r,t),
\end{equation}
where $r$ is the radial coordinate, $v_r$ is the radial component of
the velocity, $\Sigma$ is the surface mass density of the gas in the
ring, and $S(r,t)$ is a position and time dependent source term. The
angular momentum equation is \citep{Pringle1981, Rafikov2002}
\begin{equation}\label{eq:angular_momentum}
  \frac{\partial (\Sigma \, \Omega r^2)}{\partial t} 
  + \frac{1}{r}\frac{\partial (r v_r \Sigma \, \Omega r^2)}{\partial r} 
  = - \frac{1}{2\pi r} 
  \left( \frac{\partial \TTv}{\partial r} 
    - \frac{\partial \TTd}{\partial r} \right)
  + S(r,t) \, \Omega r^2 
\end{equation}
where $\TTv = -2\pi r^3 \nu \Sigma \, \partial \Omega/\partial r$ is
the viscous torque with $\nu$ the viscosity. Taking into account that
$\Omega$ (through $\Menc$) is not an explicit function of time, we
combine eq.~(\ref{eq:continuity}) and eq.~(\ref{eq:angular_momentum})
to eliminate $v_r$, so that we are left with
\begin{equation}\label{eq:time_dependent_sigma}
  \frac{\partial \Sigma}{\partial t} = 
  \frac{1}{2\pi r} \frac{\partial}{\partial r} 
  \left[ \frac{\partial (r^2\Omega)}{\partial r} \right]^{-1}
  \left( \frac{\partial \TTv}{\partial r} 
    - \frac{\partial \TTd}{\partial r} \right)
  + S(r,t).
\end{equation}
The radial motion of the star cluster is governed by the competing
torques of dynamical friction and ring tides. Henceforth, the change
of angular momentum of the star cluster with time is
\begin{equation}\label{eq:cluster_angular_momentum}
  \Ms \frac{\partial (\OFs \rs^2)}{\partial t} = \TTdf - \TTd.
\end{equation}

For the (bulge of the) spiral galaxy in which star cluster and nuclear
ring are embedded, we adopt a density profile $\rho \propto r^{-2}$
(see \S~\ref{sec:discussion_isothermal_profile} for a discussion on
the validity of such an isothermal profile). The corresponding mass
profile is given by $\Menc = \Mencz \, (r/r_0)$, with $\Mencz$ the
mass enclosed within radius $r_0$. Such a mass profile has the nice
property that the orbital velocity, $\vorb = \Omega r = \sqrt{G
  \Mencz/r_0}$, is constant, and thus $\partial (\OFs \rs^2) /
\partial t = \vorb \partial \rs / \partial t$ in
eq.~(\ref{eq:cluster_angular_momentum}). With this mass profile, the
partial derivatives of the viscous and tidal torques reduce to (see
also C08)
\begin{eqnarray}
  \label{eq:viscous torque}
  \frac{\partial \TTv}{\partial r} & = & 
  2\pi \vorb \frac{\partial (r \nu \Sigma)}{\partial r},
  \\
  \label{eq:external torque}
  \frac{\partial \TTd}{\partial r} & \simeq &  
  \beta \frac{G \Ms^2 r_0}{\Mencz} \frac{r^3 \Sigma}{(\rs-r)^4}.
\end{eqnarray}
For the viscosity we follow C08, adopting $\nu = \nu_0
(\Sigma/\Sigma_0)^2$ as suggested by \cite{Lin1986} and $\nu_0 =
\alpha (Q_0 \qd)^2/2 \rdz \vorb$ using the standard \cite{Shakura1973}
$\alpha$-disk with \cite{Toomre1964} $Q_0$ parameter, and, as before,
$\qd = \Md/\Mencz$ the ratio of the nuclear ring (gas) mass to the
enclosed mass.

Substituting these expressions and rescaling the surface mass density
and time respectively as $\sigma = \Sigma/\Sigma_0$ and $t' =
t/\tdfz$, with $\tdfz = \Mencz /(\Omega_0\Ms\ln\Lambda) $,
equations~(\ref{eq:time_dependent_sigma})
and~(\ref{eq:cluster_angular_momentum}) become
\begin{eqnarray}
  \label{eq:master1}
  \frac{\partial \sigma}{\partial t'} & = & 
  \frac{\alp}{\qp} \, \qd \frac{\rdz^2}{r} 
  \frac{\partial^2 (r\sigma^3)}{\partial r^2} 
  + \bp \qp \, \qd \frac{\rdz^3}{r} 
  \frac{\partial}{\partial r} \frac{r^3\sigma}{(\rs-r)^4} 
  + \gp \qp \, \frac{1}{\qd^{1/3}} f(r),
  \\ 
  \label{eq:master2}
  \frac{\partial \rs}{\partial t'} & = &
  \bp \, 2 \qd \rdz \int_{0}^{\rd} \frac{r^3\sigma}{(\rs-r)^4} \, \du r
  - \frac{\rdz^2}{\rs}.
\end{eqnarray}
As before, $q' = \Ms/\Md$ is the ratio of the cluster mass to the
nuclear ring (gas) mass, $\alp \equiv \alpha Q_0^2/(2\ln\Lambda)$ and
$\bp \equiv \beta/(2\pi\ln\Lambda)$ set the strength of the ring
viscous and tidal torques relative to the dynamical friction torque,
and the constant $\gp$ controls the amount of inflow torque due to gas
coming in along $X_1$ orbits that transitions to $X_2$ orbits. The
dimensionless function $f(r)$ defines the radial extent where this gas
is deposited in the nuclear ring. We choose the form
\begin{equation}\label{eq:radial_extent_inflow}
  f(r) = 
  \left\{
    \begin{array}{ll}
      \frac{1}{2\Dr}  & \mathrm{if}\ \rdz - \Dr < r < \rdz \\
      0               & \mathrm{otherwise}
      \end{array}
    \right.,
\end{equation}
so that for $\gamma = 1$, the inflow torque and dynamical friction
torque approximately cancel, corresponding to the scale of the mass
inflow rate in eq.~(\ref{eq:mass_inflow_scale}).

Equations~(\ref{eq:master1}) and~(\ref{eq:master2}) are the governing
equations governing our model. They are the same as respectively
equations~(23) and~(24) in C08, except for the extra source term at
the right-hand-side of eq.~(\ref{eq:master1}). In addition, our
initial conditions and timescales of interest are different from the
scenario studied in C08.

\section{Results}
\label{sec:results}

We first estimate the rate at which a star cluster moves out from the
nuclear ring. Next, we numerically solve the equations describing our
model, without any additional source of gas, and in case of a constant
mass inflow rate.

\subsection{Estimate of the Rate of Cluster-Ring Separation}
\label{sec:estimate_separation}

We give an order of magnitude estimate for the rate at which a newly
formed star cluster will separate itself form the nuclear ring. For
simplicity, we suppose that the gas in the nuclear ring does not react
significantly to the tidal torque, i.e., $\Sigma \simeq \Sigma_0$
($\sigma \simeq 1$) and $\rd \simeq \rdz$ remain approximately
constant. In this case, eq.~(\ref{eq:master1}) can be neglected, and
ignoring the last term in eq.~(\ref{eq:master2}) due to dynamical
friction, it simplifies to
\begin{equation}\label{eq:master2_simplified}
  \frac{\partial \rs}{\partial t'} \sim
  2 \bp \qd \rdz \int \frac{r^3}{(\rs-r)^4} \, \du r.
\end{equation}
Since $\rs-r \ll r$, the latter integral out to the edge of the ring
$\rdz$ can be approximated as $[\rdz/(\rs-\rdz)]^3/3$. This leads to
an ordinary differential equation in $x \equiv (\rs-\rdz)/\rdz$ of the
form $\du x/\du t' = (2\bp\qd/3) \, x^{-3}$. The resulting solution,
together with the Hill radius $\rHd$ defined in
eq.~(\ref{eq:hill_radius}) and $\rs \gtrsim \rdz \gg \rHd$, implies
that
\begin{equation}\label{eq:timescale_separation}
  \left( \frac{\rs-\rdz}{\rHd} \right)^4 \sim 
  \frac{8\bp}{3} \frac{\rdz}{\rHd} \frac{t}{\tdfz}
\end{equation}
Since the left-hand-side as well as $8\bp/3$ are of order unity, the
timescale for a star cluster to separate itself from the nuclear ring
in which it was formed is of order $\tsep \sim (\rHd/\rdz) \tdfz$.
The observed range of $\qd = 0.01 - 0.03$ implies $\tsep/\tdfz \sim
0.2 - 0.3$. Furthermore, in units of the crossing time $\tcr = \tdfz
\qp \qd 2\pi \ln \Lambda$, the separation time is about $\tsep/\tcr
\sim (\rHd/\rdz) /\qp$. This means a few to tens of orbits for
cluster-to-ring mass ratios from $\qp=0.01$ to $\qp=0.1$, or about
$500$\,Myr to $50$\,Myr for a typical crossing time of $\tcr =
20$\,Myr.

Since the time $t$ in eq.~(\ref{eq:timescale_separation}) goes as a
fourth power of the cluster-ring separation $\rs - \rdz$, the initial
separation is much faster than $\tsep$. This means that for a star
cluster to reach a significant fraction of the Hill radius of the
nuclear ring is very fast, i.e., typically of order less than $1\%$ of
the dynamical friction time. This implies a factor $20 - 30$ quicker
than $\tsep$, or, with $\tcr = 20$\,Myr, less than $20$\,Myr for
$\qp=0.01$ and less than $2$\,Myr for $\qp=0.1$. We come back to this
separation in the next section when we (numerically) investigate the
system in greater detail.

\subsection{Separation and Density Enhancement}
\label{sec:separation_and_density_enhancement}

We solve equations~(\ref{eq:master1}) and~(\ref{eq:master2}) using
standard explicit finite-difference methods
\citep{Press92..numrecipies}.  For good spatial resolution, we use
2000 grid points in $r/\rdz$ between zero and $1.50$, with the star
cluster initially at $\rs/\rdz = 1.01$. We set $\qd = \Md/\Menc =0.01$
similar to the observed ratios of (gas) mass in the ring to enclosed
mass within the ring radius derived in \S~\ref{sec:properties}.
Furthermore, for the dimensionless parameters $\alp$ and $\bp$,
representing the strength of the ring viscosity and tides relative to
the dynamical friction, we choose $\bp=1$ and $\alp=0.01$. We first
study the case without source term, i.e., $\gp=0$, while in the next
section \S~\ref{sec:constant_mass_inflow_rate}, we include the effect
of gas mass inflow.

\begin{figure}
  \plotone{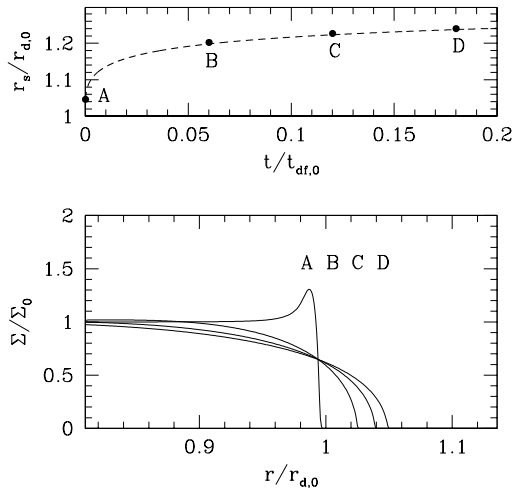}
\caption{The top panel shows the evolution of the radial position of a
  star cluster, $\rs$, relative to the initial outer radius of the
  nuclear ring, $\rdz$, as function of the time, $t$, in units of the
  initial dynamical friction time, $\tdfz$. The bottom panel shows
  snapshots of the surface density profile at the edge of the nuclear
  ring for various times, separated by $\Delta t/\tdfz = 0.06$, with
  labels A-D as indicated in the top panel. The star cluster has mass
  of $1\%$ ($\qp=0.01$) relative to the gas mass in the nuclear ring.
  Furthermore, the dimensionless model parameters are set to
  $\alp=0.01$, $\bp=1$, and $\gp=0$ (no mass inflow).}
\label{fig:md0.01}
\end{figure}

\placefigure{fig:md0.01}

In Figure~\ref{fig:md0.01}, we show the (numerical) solution of
equations~(\ref{eq:master1}) and~(\ref{eq:master2}) for a star cluster
with mass relative to the (gas) mass in the nuclear ring of $\qp =
0.01$. The top panel shows the evolution of the radial position of the
star cluster, $\rs$, relative to the initial outer radius of the
nuclear ring, $\rdz$, as function of the time, $t$, in units of the
initial dynamical friction time, $\tdfz$. The bottom panel shows
snapshots of the surface density profile at the edge of the nuclear
ring for various times, separated by $\Delta t/\tdfz = 0.06$, with
labels A-D as indicated in the top panel. The initial built-up of
material at the edge of the ring is very fast and only after $t/\tdfz
= 10^{-4}$ (point A) an enhancement of about $30\%$ in the surface
mass density is achieved just inside $\rdz$.  This rapid initial
enhancement is due to the fourth-power radial dependence of the tidal
torque in eq.~(\ref{eq:external torque}), and may not develop in a
realistic system.  Nevertheless, the initial ring tidal torque will be
very strong and pushes the star cluster outward beyond the initial
edge of the ring.  After a time $t/\tdfz = 0.06$ (point B), the star
cluster has moved out about $20\%$ from its initial position. The
surface mass density near the outer edge decreases now with respect to
its initial value due to viscous spreading. Points C and D continue
the trend of an outward moving star cluster and viscously spreading
ring, albeit at a decreasing pace.

\begin{figure}
  \plotone{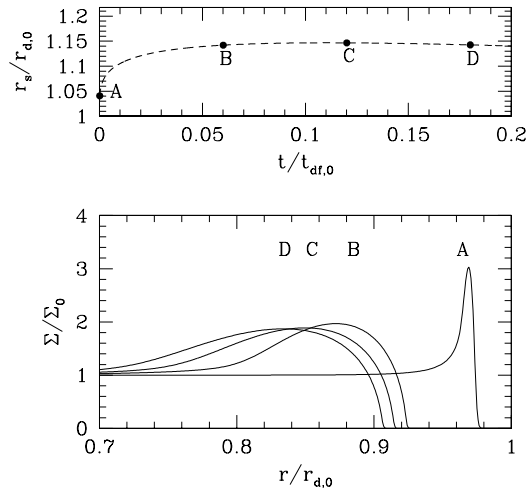}
\caption{Same as Figure \ref{fig:md0.01}, but for at ten times more
  massive star cluster with $\qp=0.1$. After an initial outward
  migration the star clusters turns around and starts moving inwards.
  At the same time, the surface mass density is pushed to smaller
  radii and enhanced by a factor three to two.}
\label{fig:md0.1}
\end{figure}

\placefigure{fig:md0.1}

In contrast, for the case of $\qp=0.1$ (a ten times more massive star
cluster), the star cluster does not continuously move outward, but
rather stop and even moves inward as shown in Figure~\ref{fig:md0.1}.
As in the earlier case, an initial surface mass density enhancement is
built at point A, though it is a factor of ten larger than for the
case $\qp=0.01$, i.e., $300\%$ ($\qp=0.1$) as opposed to $30\%$
($\qp=0.01$) of its initial value.  However, the subsequent evolution
of this case differs from the earlier case. The star cluster still
moves outward, but it reaches only about $15\%$ of its initial
position. Its maximum in $\rs/\rdz$ is reached between point B and C,
after which the star cluster ``turns around'' and starts moving inward
with respect to the initial outer radius of the ring. The surface mass
density at the edge of the ring monotonically moves inward,
effectively being ``shepherd'' inward by the star cluster, similar to
the scenario in C08.

\begin{figure}
  \plotone{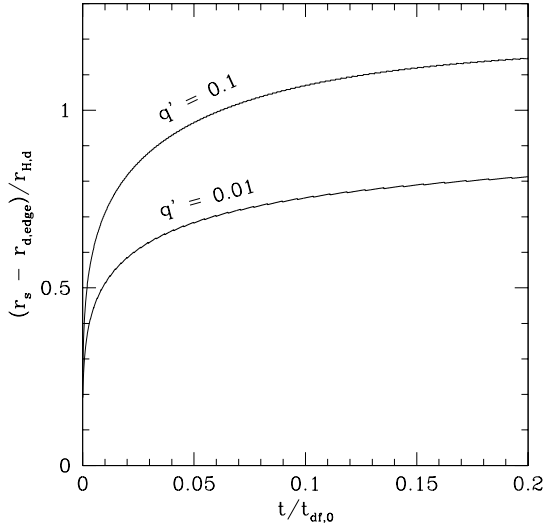}
\caption{The evolution of the radial separation $\rs-\rde$ between
  the star cluster and the edge of the nuclear ring, normalized by the
  Hill radius $\rHd$ of the nuclear ring. The top curve is for a ten
  times more massive star cluster ($\qp=0.1$, see
  Figure~\ref{fig:md0.1}) then bottom curve ($\qp=0.01$, see
  Figure~\ref{fig:md0.01}). For both cases, the time to converge to
  the final separation is about $20\%$ of the initial dynamical
  friction time $\tdfz$, while the initial separation is very fast,
  reaching half of the final separation in less than $1\%$ of
  $\tdfz$.}
\label{fig:ring_t_vs_xp}
\end{figure}

\placefigure{fig:ring_t_vs_xp}

In both cases, the radial separation between the star cluster and the
(edge of the) nuclear ring approach a constant value, which we expect
from the order of magnitude estimate in \S~\ref{sec:basic_picture} to
be close to the Hill radius of the ring, $\rHd$, as defined in
eq.~(\ref{eq:hill_radius}). Substituting $\qd = 0.01$ and using that
$\rs \gtrsim \rdz \gg \rHd$, we estimate this Hill radius to be around
one-fifth of the initial radius of the ring, i.e., $\rHd/\rdz \sim
0.2$. Combining the above shifts in the radial positions of the star
cluster and nuclear ring, we find that the separations for both cases
indeed converge to around one-fifth of $\rdz$, although for $\qp=0.01$
the separation stays slightly smaller, while for $\qp=0.1$ it is
somewhat larger.
This is further illustrated in Figure~\ref{fig:ring_t_vs_xp}, where we
show the evolution of the radial separation $\rs-\rde$ normalized by
the Hill radius $\rHd$. We use two definitions of the radius of the
disk edge, $\rde$. For $\qp = 0.01$, we define $\rde$ as largest $\rd$
for which $\Sigma/\Sigma_0>75\%$, while for $\qp = 0.1$, $\rde$ is
defined as the $\rd$ for which $\Sigma$ peaks.  Despite these two
different definitions, Figure~\ref{fig:ring_t_vs_xp} shows two
important points. First, the separation between the disk and satellite
is of order $\rHd$, consistent with our expectations from
\S~\ref{sec:basic_picture}.  Second, the final separation is reached
after about $20\%$ of the dynamical friction time, consistent with the
estimate of $\tsep/\tdfz \sim \rHd/\rdz \sim 0.2$ in
\S~\ref{sec:estimate_separation} for $\qd=0.01$. This figure also
confirms that the initial separation is very quick, indeed reaching
half of the final separation within a one percent of the dynamical
friction time, also consistent with our earlier expectations.

\subsection{Constant Mass Inflow Rate}
\label{sec:constant_mass_inflow_rate}

\begin{figure}
  \plotone{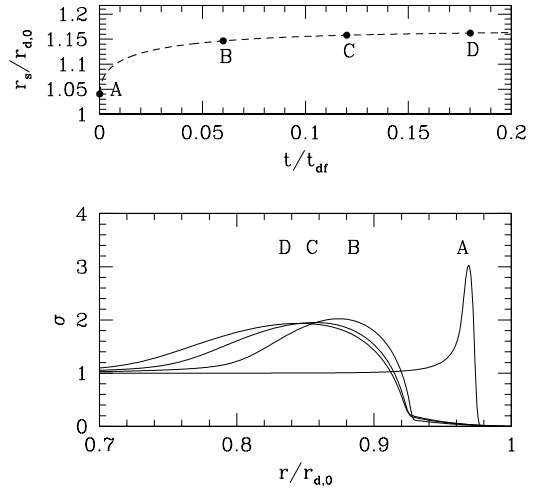}
\caption{Same as Figure~\ref{fig:md0.1} with $\qp=0.1$, but now with
  $\gp=1$. The torque due to the latter mass inflow rate approximately
  cancels the dynamical friction torque or the star cluster,
  preventing its radius $\rs$ to shrink. The long tail extending from
  $r/\rdz \sim 0.92$ out to $r/\rdz \sim 1$ is due to mass inflow into
  this region.}
\label{fig:md0.1_mdot0.1}
\end{figure}

We now include the effect of a constant gas mass inflow on the
dynamics of the cluster-ring systems. Figure~\ref{fig:md0.1_mdot0.1}
shows the results of (numerically) solving the same system with
$\qp=0.1$ as in Figure~\ref{fig:md0.1}, except that the source term in
eq.~(\ref{eq:master1}) now contributes with $\gp=1$. We estimated in
\S~\ref{sec:estimate_separation} that the latter corresponds to an
mass inflow rate of $\dot{M} \sim 0.1$\,\Msunyr. 
Comparing the top panels of the two figures, we find that the star
cluster moves (slightly) further out when $\gp = 1$ and remains
constant at the maximum radial position, as opposed to the shrinking
of $\rs$ after having moved out for an initial period when $\gp = 0$.
This is because in the case that $\gp = 1$, the inflow torque is
balancing the torque due to dynamical friction, as we estimated in
eq.~(\ref{eq:mass_inflow_scale}).
The bottom panels of the two figures are very similar, both showing an
enhancement as well as inward migration of the surface mass density.
The most significant difference is that in
Figure~\ref{fig:md0.1_mdot0.1} beyond the edge at $r/\rdz \sim 0.92$
there is a long tail out to $r/\rdz \sim 1$, due to mass inflow into
this region. Another difference is that for $\gp=1$ after the initial
inward push of the edge, it remains fixed at $r/\rdz \sim 0.92$,
whereas for $\gp=0$ it still moves inward, although at a decreasing
rate.  This is consistent with the above difference in the change in
the radial position $\rs$ of the star cluster. As a consequence, also
in the case that $\gp=1$, the separation between the star cluster and
the (edge of the) nuclear ring converge to a constant value, which
is again of the order of the Hill radius of the ring.

\placefigure{fig:md0.1_mdot0.1}

\begin{figure}
  \plotone{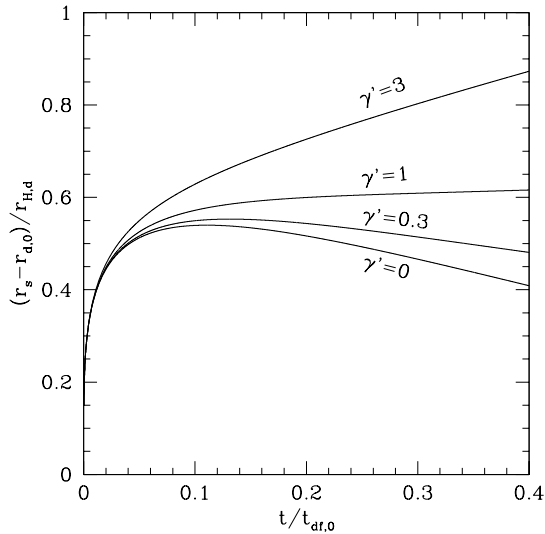}
\caption{
  The evolution of the radial position of a star cluster with
  $\qp=0.1$, relative to the initial outer edge of the ring,
  $\rs-\rdz$, in units of the Hill radius of the ring, $\rHd$, for
  four different constant mass inflow rates $\gp = 0$, $0.3$, $1$ and
  $3$ from bottom to top. After a similar initial outward migration,
  the motion of the star cluster goes from in-falling for $\gp \lesssim
  1$ to outgoing for $\gp \gtrsim 1$.}
\label{fig:ring_t_vs_xp_mdot}
\end{figure}

In Figure~\ref{fig:ring_t_vs_xp_mdot}, we show the evolution of the
radial position of the star cluster relative to the initial outer edge
of the ring $\rs-\rdz$, in units of the latter Hill radius $\rHd$, for
four different constant mass inflow rates. From bottom to top the
curves correspond to $\gp = 0$, $0.3$, $1$ and $3$. This clearly shows
the transition around $\gp = 1$: after a similar initial outward
migration, the motion of the star cluster goes from in-falling for $\gp
\lesssim 1$ to outgoing for $\gp \gtrsim 1$. This is fully consistent
with our estimated scale for the mass inflow rate $\dot{M}$ in
eq.~(\ref{eq:mass_inflow_scale}), case of single star cluster with
$\qp=0.1$.

\placefigure{fig:ring_t_vs_xp_mdot}

\section{Comparison to Observations}
\label{sec:comparison}

We describe the expected evolution of nuclear rings due to interaction
with star clusters based on our findings in the previous sections.
First, we focus on the evolution in radius, and discuss possible
azimuthal variations thereafter. Finally, we take a more detailed look
at the well-studied nuclear ring in NGC\,4314.

\subsection{Radial Evolution of Nuclear Rings}
\label{sec:radial_evolution}

From our analysis in \S~\ref{sec:tidal}, we expect that a newly formed
star cluster moves outward until it is about a Hill radius separated
from (the edge of) the nuclear ring. We derived in
\S~\ref{sec:basic_picture} a range in $\rHd$ from about $50$\,pc to
$500$\,pc, with $150$ pc as a typical value. For a host galaxy at
distance of $\sim 10$\,Mpc, this corresponds to about $1\arcsec$ to
$10\arcsec$, and typically $3\arcsec$, which should be readily
observable. However, while the position of the (young) star clusters
with respect to the center of their host galaxy is straightforward to
measure, the radial extent of the gas in the ring is often difficult
to determine. It requires high spatial resolution CO measurements to
trace the underlying molecular gas. Also, part of the gas (at the
edge) of the ring may be severely disturbed due to winds that clear
out the gas and dust from the newly formed star clusters and/or
ionized by the massive, bright stars in the young star cluster (but
see \S~\ref{sec:discussion_star_formation} and
\S~\ref{sec:discussion_stellar_feedback}). Still, the dust seems to be
well mixed into the gas \citep[e.g.][]{Brunner2008}, indicating that
dust provides a good tracer of the extent of the gas in the ring.
Apart from the contact points where the bar dust lanes connect with
the nuclear ring, the dust is indeed found on the inside of the
distribution of the star clusters (see also
\S~\ref{sec:basic_picture}).

Still, using the (peak in the) dust distribution as an estimate of the
edge of the ring, the (unique) measurement of a separation is
complicated by the spread in radius of the star clusters. Part of this
spread is because of the dependence of the dynamical friction time on
the mass of the star cluster: $\tdf/\tcr \simeq 1/\qp$. Even though we
found that it takes less than $1\%$ of the dynamical friction time to
reach about half of the Hill radius, with a typical crossing time of
$\tcr = 20$\,Myr, this means only $2$\,Myr for the most massive star
clusters ($\qp=0.1$), but as much as $200$\,Myr for the least massive
clusters ($\qp=0.001$). Since $\tsep/\tdf \sim 0.2 - 0.3$, to reach
the full separation takes a factor $20 - 30$ longer. This means that
only the most massive clusters reach at least half of the Hill radius
separation before all bright, massive stars die. The less massive
clusters have already faded significantly, or even dissolved. So, we
then might expect a smooth radial gradient with the most massive
clusters on the outside, which in turn can be used to establish the
Hill radius.

Unfortunately, we found that the more massive cluster indeed rapidly
reach a maximum separation, but after staying a while at this radius
(depending on the mass inflow rate), they move inward again by pushing
the ring inward (compare for example the top panels of
Figures~\ref{fig:md0.01} and~\ref{fig:md0.1}).  Hence, over time the
less massive clusters, if they survive, may catch up with the more
massive clusters. We conclude that only among the (very) young (few
Myr) clusters we might expect a radial gradient with the more massive
clusters to the outside, from which the Hill radius can be estimated.
Later on, if a radial gradient is visible at all, it might even be
opposite, with the more massive clusters on the inside.  Since the
more massive clusters have higher survival chance, the latter 'turn
around' may result in at most a mild radial gradient in age (with the
older clusters on the inside). \cite{Mazzuca2008} find only for 2 (out
of 22) nuclear ring galaxies, a radial age gradient with indeed
(slightly) older clusters on the inside.

\subsection{Azimuthal Age Gradient}
\label{sec:aziumthal_age_gradient}

Unlike the radial age gradient, \cite{Mazzuca2008}, as in other
detailed studies \citep[e.g.\ by][]{Boker2008}, do find signatures of
an azimuthal age gradient in most of the nuclear rings, ranging from
clean bipolar age gradients to a correlation between the youngest star
clusters and the contact points of the bar dust arms with the nuclear
ring. The natural explanation is that at (or very close to) these
contact points the inflowing gas shocks and forms stars, after which
the resulting star cluster travels in the ring away from the contact
points. Indeed, for a typical bar pattern speed of $\sim 50$\,\kmskpc\ 
\citep[e.g.][]{Gerssen2002} the corresponding time for the contacts
points to rotate is $\sim 120$\,Myr, significantly shorter than the
local crossing time $\tcr \sim 20$\,Myr. Alternatively, if the star
formation occurs due to (local) gravitational instability of the gas
in the ring, the star clusters are expected to form throughout the
ring without an azimuthal age gradient. Instead of associating this
instability with an inner Lindblad resonance \citep[ILR;
e.g.][]{Elmegreen1994}, it is a natural consequence of shepherding:
The gas in the ring loses angular momentum to the star cluster(s),
which in turn lose angular momentum to the host galaxy through
dynamical friction.  This means that gas at the edge of the ring moves
inward, resulting in an increase in the gas surface density at the
edge (see also Figures~\ref{fig:md0.01}, \ref{fig:md0.1}
and~\ref{fig:md0.1_mdot0.1}), until its becomes unstable against
self-gravity.

In the case of shepherding we do expect, as observed, a mixture in
azimuthal age gradients. Initially, when the gas surface density in
the ring is below the critical value for star formation, star
clusters only form in episodes at the two contact points, they move
together with the remaining gas along the ring, resulting in bipolar
age gradient.  The newly formed star clusters move outward, while at
the same time 'pushing' the gas inward, causing an over-density
built-up at the edge of the ring. When the gas surface density
somewhere along the edge of the ring reaches the critical value, it
becomes unstable and forms stars. The more massive the star cluster
(or group of star clusters acting together), the higher the
over-density and the sooner the critical value is reached. Generally,
the most massive clusters and hence the star formation might be
anywhere along the ring. However, depending on the mass inflow rate,
there might be relatively more gas near the contact points to start
with, so that first instability-driven star clusters still form near
the contact points.  Therefore, an azimuthal age gradient might still
exist over part of the ring, but eventually the gradient is expected
to flatten.

We found in \S~\ref{sec:constant_mass_inflow_rate} that for a low
enough mass inflow rate, the (most massive of the) existing star
clusters move inward by dynamical friction, while 'pushing' and
increasing the surface mass density of the gas in the outer part of
the nuclear ring. As a result, the gas might reach the critical
surface mass density and form new stars (in clusters) all along the
edge of the ring. At the same time, near the contact points, which
also move inward along with the nuclear ring, star formation from
inflowing and shocking gas might be ongoing. We argue below in
\S~\ref{sec:ngc4314} that the nuclear ring in NGC\,4314 is an example
of this mechanism.  Since even in the case of a high mass inflow rate,
eventually enough (massive) star clusters are formed for their
(combined) dynamical friction to 'push' to ring inward, the
corresponding gas and bright star clusters migrate inward from larger
to smaller $X_2$ orbits. In other words, nuclear rings will be located
between the so-called outer and inner ILR, although strictly speaking
resonances are only applicable to systems with a weak bar. Depending
on the details of the mass inflow rate and star formation efficiency
over time, we thus expect a variety in sizes of nuclear rings with
respect to the resonance radii, as well as a significant diversity in
their star cluster distribution. Hydrodynamic simulations indeed show
the inward migration of nuclear rings
\citep[e.g.][]{Regan2003,Fukuda2000}, and observations of nuclear
rings confirm the diversity in their star cluster distribution
\cite[e.g.][]{Mazzuca2008}

\subsection{Nuclear Ring in NGC\,4314}
\label{sec:ngc4314}

NGC\,4314 is a nearby ($D \sim 10$\,Mpc, $1\arcsec \simeq 48.5$\,pc)
galaxy with a large-scale bar out to a radius of $\simeq 65\arcsec$
(3.15\,kpc). It host a prominent circumnuclear ring of star formation
that peaks at a radius of $\simeq 7.0\arcsec$ (340\,pc), that is
visible in \Ha, radio continuum as well as optical color maps.  High
spatial resolution CO observations by \cite{Benedict1996} clearly show
that the molecular gas peaks at a smaller radius of $\simeq
5.5\arcsec$ (270\,pc).

The circular velocity curve flattens at $\sim 7\arcsec$ to a value of
$\sim 190$\,\kms\ \citep{Combes1992}, corresponding to a (spherically)
enclosed mass $\Menc \sim 2.8 \times 10^9$\,\Msun.  The total
molecular gas mass within $7\arcsec$ is $\sim 15 \times 10^7$\,\Msun.
However, this includes gas that is still flowing in along the dust
lanes at the contact points, as well as a significant amount in the
center where a strong peak in the CO coincides with a peak in
$\Ha+[\NII]$ consistent with the LINER definition. Hence, we estimate
the ring-to-enclosed mass to be $\qd \sim 0.03$ \citep[see
also][]{Benedict1996}. The Hill radius of the ring is thus $r_H \sim
105$\,pc or $r_H \sim 2.2\arcsec$. The observed difference of $\simeq
1.5\arcsec$ between the peaks in star formation and CO flux is smaller
than the predicted Hill radius, which is expected based on the
discussion in \S~\ref{sec:radial_evolution}.
Although the peak in CO flux nicely corresponds with the peak in the
dust distribution and thus likely traces the edge of the ring, the
star clusters have not yet reached the Hill radius. Given a separation
time of $\tsep/\tdf \sim \qd^{1/3} \sim 0.3$, a dynamical friction
time of $\tdf/\tcr \simeq 1/q'$, a crossing time of $\tcr \sim
11$\,Myr, and an average age of the star clusters of $t \sim 15$\,Myr,
we expect based on eq.~(\ref{eq:timescale_separation}), that the star
clusters have reached a separation of about $\Dr \sim 1.5 \, \qp^{1/4}
\, \rHd$. For star clusters with a (combined) mass relative to the gas
mass in the ring from $\qp=0.01$ to $\qp=0.1$, this means a separation
from about $1.0\arcsec$ to $1.8\arcsec$, bracketing the observed
difference.

There is no evidence for an azimuthal age gradient, but a slight
tendency for a radial age gradient with the few clusters outside the
ring begin all older than $15$\,Myr. This is consistent with the
shepherding scenario in which the gas in the ring has been 'pushed'
inward until a critical surface density is reached and all along the
edge of the nuclear ring star clusters are formed, which then move
outward.  The (group of) star cluster(s) responsible for the inward
migration of the ring likely have faded too much to be visible against
the old underlying stellar population of the host galaxy. However,
further support for the inward migration of the ring comes from two
blue stellar spiral arms that seems to connect the outer-most $X_2$
orbits (also referred to as the outer ILR) with the nuclear ring
\citep{Wozniak1995, Benedict2002}. Subtracting a model of the red host
galaxy, \cite{Benedict2002} find that the stars in the arm close to
the nuclear ring are $\sim 20$\,Myr old, with an indication of an
increase in age beyond $\sim 100$\,Myr going outward in the arms.
Whereas during the inward migration the gas surface density along
(the edge of) the ring was insufficient to form stars, shocks at the
contact points might still have allowed the formation of stars. While
the contact points moved inward together with the ring, these stars
stayed behind and created the blue arms.

\cite{Benedict1996} derive an outer ILR at $\sim 13\arcsec$ and inner
ILR at $\sim 4\arcsec$ for an assumed bar pattern speed $\Omega_p \sim
72$\,\kmskpc\ of the large-scale bar. The latter is likely an upper
limit because typically the co-rotation radius is beyond the $\sim
3$\,kpc radius of the bar, so that with a $190$\,\kms\ circular
velocity, $\Omega_p \lesssim 60$\,\kmskpc. In turn this implies that
the above outer and inner ILR are probably lower and upper limits
respectively. This clearly shows that it is critical to obtain an
independent measurement of the bar pattern speed, using for example
the Tremaine-Weinberg (\citeyear{Tremaine1984}) method. Even so, the
nuclear ring is in between the outer and inner ILR, in line with the
shepherding scenario, in which the ring migrates inward until the
increasing gas surface density at the edge reaches the critical value
for star formation. While the newly formed star clusters move
(slightly) outward peaking at $\sim 7.0\arcsec$, the gas in the ring
is 'pushed' further inward peaking at $\sim 5.5\arcsec$. Note that
\cite{Benedict1993} detected a nuclear bar within $\sim 4\arcsec$ that
might provide a mechanism to transport part of the gas further inward
towards the presumed black hole in the center \citep[see
also][]{Benedict2002}.

\section{Discussion}
\label{sec:discussion}

We discuss some of the assumptions of our model and we speculate about
possible consequences the inward migration of the cluster-ring system
has on transporting gas to the center of the galaxy.

\subsection{Isothermal Density Profile}
\label{sec:discussion_isothermal_profile}

In deriving the equations for our model in
\S~\ref{sec:basic_equations}, we have adopted a density profile $\rho
\propto r^{-2}$ for the (bulge of the) spiral galaxy in which the star
cluster and nuclear ring are embedded. At the radii where nuclear
rings are present \citep[0.2--1.7\,kpc;][]{Mazzuca2008}, the density
of galaxies indeed seem to be not too far from such an isothermal
profile, as indicated, for example, by studies that combine stellar
dynamics with strong gravitational lensing
\citep[e.g.][]{Koopmans2006, vandeVen2008}. A more straightforward way
is to consider that the surface brightness profiles of (bulges of)
galaxies are well fitted by a \cite{Sersic1968} profile $I(R) \propto
\exp[-(R/R_e)^{1/n}]$, with index $n$ and the effective radius $R_e$
enclosing half of the total light. The (numerically) deprojection is
well approximated by the density profile of \cite{Prugniel1997}, of
which the (negative) slope is given by
\begin{equation}\label{eq:deprojected_Sersic_slope}
  \gamma = p_n + \frac{b_n}{n} \left(\frac{r}{R_e}\right)^{1/n},
\end{equation}
with $p_n=1.0-0.6097/n+0.05463/n^2$ and $b_n = 2\,n - 1/3 + 4/405\,n +
46/25515\,n^2$ \citep{Ciotti1999}. For an average bulge of a spiral
galaxy $n \sim 2$ and $R_e \sim 1$\,kpc \citep[e.g.][]{Hunt2004}, so
that at the typical radius $r \sim 0.5$\,kpc of a nuclear ring, we
indeed find that the slope of the density $\gamma \sim 2$ is close to
isothermal.

\subsection{Gas Clumps and Clouds}
\label{sec:discussion_gas}

The star cluster that is being formed in the nuclear ring is already
subject to dynamical friction while it is still a gas clump, as long
as the material that makes up the (proto) star cluster is bound enough
relative to the background stellar population of the host galaxy.
Since this gas clump contains (significant) more mass than the final
cluster of only stars, the initial dynamical friction is (much)
stronger. However, the clearing of gas (and dust) from the newly
formed star cluster is very fast and efficient, since even the
youngest visible star clusters are only mildly reddened
\cite[e.g.][]{Maoz2001}.

We assume that the gas in the nuclear ring is uniform, i.e., no gas
clouds (or clumps) aside from those which turn into star clusters.
Since such gas clouds would also be subject to dynamical friction from
the background bulge stars and thus move inward, dynamical friction on
the nuclear ring itself has also been proposed as an explanation of
the inward migration of the nuclear ring in for example NGC\,4314
\citep{Combes1992}.

Suppose that the gas is made up of $N$ gas clouds of the same mass
$\Mc = \Md/N$, then about $N = 1/\qp^2$ clouds are needed to have a
combined dynamical friction torque of the same order as that of a
single star cluster of mass $\Ms = \qp \Md$. Adopting as before the
two cases of $\qp=0.01$ and $\qp=0.1$, this respectively means
$N=10^4$ and $N=10^2$ gas clouds of mass $\Mc = 5 \times 10^3$\,\Msun\
and $\Mc = 5 \times 10^5$\,\Msun\ for a typical nuclear ring mass of
$\Md = 5 \times 10^7$\,\Msun. 
From \S~\ref{sec:properties} we have that the dynamical friction time
in units of the crossing time is approximately given by $\tdf/\tcr
\simeq \Md/\Ms$, so that with $\Ms = \Mc = \Md/N$ for a gas cloud we
obtain $\tdf \simeq N \tcr$. Given a typical crossing time of $\tcr
\sim 20$\,Myr, the timescales of dynamical friction for the latter two
cases are about $\tdf \sim 200$\,Gyr and $\tdf \sim 2$\,Gyr,
respectively.
We thus conclude either nearly no inward migration ($\qp \sim 0.01$),
or gas that is very clumpy with quite large gas clouds ($\qp \sim
0.1$).

Moreover, since the gas clouds will span a range in mass with
different corresponding dynamical friction timescales, they migrate
inward at different rates, leading to a (radial) diffusion of the
nuclear ring, especially with respect to the gas that is not in clouds
and stays behind. Although in some galaxies, like M\,83, that are
claimed to harbor a (partial) nuclear ring, the corresponding
morphology is rather clumpy and chaotic, most of the nuclear rings are
are quite smooth and well defined, including the nuclear ring in
NGC\,4314. Collisionless star clusters are also less prone to
destruction than collisional gas clouds, in particular in the presence
of shear stresses from differential speeds. As a result, at least the
most massive star clusters will sink faster inward than the gas
clouds, and appear to be on the inside of the gas (clouds) in the
nuclear ring, which typically is \emph{not} observed.

\subsection{Star Formation within the Nuclear Ring}
\label{sec:discussion_star_formation}

We have assumed that the star cluster form just exterior of the
nuclear ring. This assumption simplifies the physics and the numerical
computation, though it is not absolutely necessary. If the star
cluster forms \emph{near} the outer edge of the nuclear ring, we do
not expect the resulting dynamics to be significantly different. By
near, we demand that the separation between the star cluster and the
outer edge is within one Hill radius of the nuclear ring. In this way,
the amount of mass which participates in the tidal evolution of the
cluster-ring system \emph{exterior} to the star cluster is less then
\emph{interior} to the star cluster. As long as this condition is
fulfilled, the evolution should proceed along the lines of our
calculation.

There are a few reasons to believe that the star clusters (preferably)
form near the outer edge of the nuclear ring. The supply of gas from
the bar merges with the ring at the outermost radius. It is unknown
how the gas spreads radially over the rest of the ring, but if it is
due to viscous processes, for instance the magneto radial instability
\citep[MRI;][]{Balbus1991}, then the surface density is maximal at the
outer edge. As a result, the gas in the ring is most likely to
be(come) unstable near the outer edge and hence preferentially
fragment there. 

While many clusters will form near the outer edge (and hence
susceptible to shepherding) for the reasons given above, clusters are
also likely to form inside of the outer edge.  For instance, once
massive star clusters are formed, their interaction with the gas ring
may result in large and fairly symmetric enhancements in the gas
surface mass density (see for example Figure~\ref{fig:md0.1}), so that
new, but significant smaller, star clusters can be formed
\emph{around} these enhancements. For the relatively small star
clusters formed on the inside, we expect the inward migration to
follow that of forming proto planets \citep[see e.g.][]{Ward1997}.
In addition, if the gas flowing along the bar is very clumpy, possess
a different vertical scale height, or a different inclination than the
gas ring, it may not (only) merge at the outermost radius. In this
way, even rather large star clusters might form on the inner edge of
the nuclear ring. They then migrate further inward (relatively faster
than cluster formed on the outer edge as they encounter less of the
gas ring), possibly along an inward extension of the inspiraling gas.
Whereas these are possible explanations for the observation that some
star clusters appear more on the inside of the nuclear ring (see also
\S~\ref{sec:basic_picture}), alternatively they might form already
inside of the ring as the result of the local interaction of a
non-axisymmetric structure such as a nuclear bar or spiral with the
interstellar medium (see also \S~\ref{sec:discussion_feeding_nucleus}).

\subsection{Stellar Feedback}
\label{sec:discussion_stellar_feedback}

Winds originating from (massive) stars and/or supernovae in the star
clusters might not only clear out the remaining gas in the star
cluster itself, but also push away part of the gas in the nuclear
ring. This creates a separation between the cluster and the ring that
might mimic the outward migration of the star cluster. However,
since the density of the interstellar medium is far less than that of
the gas in the ring, the winds will escape in the vertical direction
like a fountain arising from the (equatorial) plane of the ring.
Therefore, even if the cluster is embedded in the ring itself, the
winds are expected to clear out at most a region that is equal to
scale height of the disk, $\hd$, which is significantly smaller than
the Hill radius of the ring, $\rHd$. 

To show this, we start from \cite{Toomre1964} $Q$ parameter for the
gas in the ring: $Q = c_s \kappa/\pi G \Sigma$ \citep{BT87}. Here, the
epicycle frequency $\kappa$ is defined as $\kappa^2 = (4 +
\du\ln\Omega/\du\ln r) \Omega^2$, so that $\kappa = \sqrt{2} \Omega$
for the assumed (isothermal) density profile $\rho \propto r^{-2}$.
Adopting $c_s = \hd \Omega$, we obtain for the scale height of the
ring $\hd/\rd \simeq (\Md/\Menc) \, Q/\sqrt{2}$. On the other hand,
the Hill radius of the ring $\rHd/\rd \simeq (\Md/\Menc)^{1/3}$.
Assuming $Q \simeq 1$ and substituting the mean observed
ring-to-enclosed mass ratio $\Md/\Menc \simeq 0.02$, we find that
$\hd/\rHd \simeq 0.05$. This indeed shows that winds might clear out a
region that is at most a few per cent of the Hill radius of the
nuclear ring.

A second possible stellar feedback effect, is due the (ultraviolet)
radiation coming from the massive, bright stars in the (young) star
cluster, which might (photo)ionize the (molecular) gas in the nuclear
ring.  We can estimate the radial extent of this ionization from the
radius $\rion$ of the Str\"omgren sphere.
We start from $\dot{N}_\mathrm{ion} \simeq \alpha \, n_g^2 \, 4\pi
\rion^3/3$, with cross section $\alpha \simeq 3 \times
10^{-13}$\,\cms. We assume that bright stars radiate roughly at $10\%$
of the Eddington luminosity in the ultraviolet (UV; $\ge 10$\,eV), to
obtain an ionization rate of $\dot{N} < 10^{48} \,
(\Mstar/\mathrm{M}_\odot)$ photons per second. With the above estimate
of the scale height $\hd$, we find for the gas mass density $\rho_g
\simeq \Menc\sqrt{2}/\pi \rd^3 Q$.  Substituting $\Menc \sim 2.5
\times 10^9$\,\Msun, $\rd \sim 0.5$\,kpc and $Q \simeq 1$, and
dividing by the proton mass, we find for the number density of the gas
$n_g \sim 4 \times 10^2$\,cm$^{-3}$. In this way, we estimate that a
(UV) radiating star of mass $\Mstar$ is able to ionize gas out to a
(maximum) radius of $\rion < 3 \,
(\Mstar/\mathrm{M}_\odot)^{1/3}$\,pc.

For a typical O star of $\Mstar \sim 40$\,\Msun, this means $\rion <
10$\,pc, which is significantly smaller than the typical Hill radius
of $\rHd \sim 150$\,pc. The combined (and focused) ionizing radiation
of thousands of massive young stars with a lifetime of only a few to
ten Myr are needed to reach the latter radius. Even then, the dust,
which is hardly effected by the radiation, will still provide a good
tracer of the gas in the ring, and as such can be used to establish
the separation between star clusters and the nuclear ring.

\subsection{Feeding the Nucleus?}
\label{sec:discussion_feeding_nucleus}

For sufficiently massive star clusters and/or for a sufficiently large
number of them, it is in principle possible for these star clusters to
shepherd the nuclear ring to smaller radii. As in C08, we assumed the
gas ring sits in a simple spherically symmetric potential. However,
the gas ring arises as a result of the stellar bar, which is a
non-axisymmetric feature in the potential, driving the gas inward
towards the $X_1-X_2$ orbit transition. In addition, the gas ring is
self-gravitating with Toomre's (\citeyear{Toomre1964}) $Q$ near unity.
We now speculate on the effect of this non-axisymmetry and self-gravity
on the gas shepherding scenario.

First, as the gas ring is being shepherded inward, it moves from one
$X_2$ orbit to a smaller one.  Eventually, it may move inward of the
smallest non-intersecting $X_2$ orbit in a barred potential
\citep[e.g.][]{Regan2003}. The gas which is pushed beyond this point
will find itself in intersecting orbits, losing angular momentum on
each orbit as it shocks against gas on the same orbit or on
neighboring orbits. The high(er) density regions resulting from the
shocks, might give rise to the observed (chaotic) spurs of gas and
dust in the centers of (late-type) galaxies
\citep[e.g.][]{Elmegreen1998,Martini2003a}.

Second, the effect of self-gravity in the shepherded gas ring is
nontrivial.  As we stated above in
\S~\ref{sec:discussion_star_formation}, the effect of an enhanced
surface density at the edge of the ring results in $Q \lesssim 1$, if
the initial $Q \simeq 1$. In addition, this \textit{local} enhancement
in the surface density may give rise to the gaseous analogy of stellar
self-gravitating groove and edge modes \citep[see
also][]{Lovelace1978, Sellwood1991, Papaloizou1991}. This may result
in the formation of nuclear bars \citep[e.g.][]{Laine2002, Erwin2002,
  Englmaier2004} or nuclear spirals \citep[e.g.][]{Englmaier2000,
  Pogge2002}, providing a means for inward transport of gas from the
nuclear ring region in analogy to the ``bars within bars'' scenario
\citep{Shlosman1990}.

\section{Conclusions}
\label{sec:conclusions}

We have studied the dynamics between a star cluster and the gas in the
nuclear ring from which it was formed. The star cluster is subject to
dynamical friction with the background bulge stars in the host galaxy. The
resulting dynamical friction torque induces a tidal (and viscous)
torque on the gas in the nuclear ring. In addition, we investigated a
torque due to in-falling gas along the bar onto the nuclear ring at the
contact points. We arrived at the following, observationally testable
predictions based on our model.

\begin{enumerate}
\item[(i)] The star clusters which are formed near the outer edge of
  the ring migrate outward. The final separation is of the order of
  the Hill radius of the ring, which is $20-30\%$ of its radius if the
  ring's gas mass is the observed $1-3\%$ of the enclosed mass.
  Similarly, the time to reach this final separation is about
  $20-30\%$ of the dynamical friction time. The initial separation is
  very fast, reaching half the Hill radius within only $1\%$ of the
  dynamical friction time. The latter can be as short as a few million
  years for a massive enough ($\gtrsim 10^6$\,\Msun) star cluster.
\item[(ii)] While such a massive cluster moves out the surface mass
  density of the gas at the edge of the nuclear ring is pushed inward
  and gets enhanced by a factor of a few. If this enhancement reaches
  the critical value for star formation, new star clusters are
  expected to form along the ring. This is in addition to the
  (ongoing) formation of star cluster from shocking gas near the
  contact points.  The (initial) azimuthal age gradient that is
  expected from the latter gets diluted or even washed out by the
  former (secondary) star cluster formation.
\item[(iii)] If the (formed) star cluster are massive and/or numerous
  enough, and the gas mass inflow rate low enough (or even shut down),
  we expect the inward migration of the cluster-ring system as a
  whole.  This mean that, even though initially the nuclear ring could
  have formed as often suggested at an (inner) Lindblad resonance, it
  can migrate closer to the center of the host galaxy.
\end{enumerate}

The latter implies that the most massive star cluster, after an
initial outward separation, turn around and migrate inward along with
the nuclear ring. Consequently, any initial radial age gradient due to
the dynamical friction time being inverse proportional to the mass of
the star cluster, is diluted or even inverted. Even so, both a radial
and azimuthal age gradient are observationally difficult to measure.
When the most massive stars are still alive, the bright emission lines
can be used for (relative) age dating \citep[e.g.][]{Boker2008}, but
after a few tens of million years it becomes already very challenging
to distinguish the fading star clusters from the background stellar
population \citep[but see e.g.][]{Benedict2002}.

The outward migration of a least the most massive star clusters is
easier to establish, and using the dust to trace the gas, observations
indeed show the young star clusters exterior to the nuclear ring. The
measured separation should be of the order of the Hill radius of the
nuclear ring, which indeed is the case in NGC\,4314. This well-studied
nuclear ring also seems to have migrated inward, leaving a tail of
older star clusters behind and being currently located between the
inner and outer ILR. To firmly confirm such a ``shepherding of gas''
towards smaller radii, the dynamics of the host galaxy, including the
(bar) pattern speed, needs to be measured accurately.


\acknowledgements

We thank Torsten B\"oker, Jes\'us Falc\'on--Barroso, Norm Murray,
Eliot Quataert and Scott Tremaine for useful discussions. We thank
Nathan Smith for pointing out the importance of stellar winds and
stellar radiation on the properties of the nuclear gas ring. We thank
the referee, Curt Struck, for insightful comments on this work.  GvdV
acknowledges support provided by NASA through Hubble Fellowship grant
HST-HF-01202.01-A awarded by the Space Telescope Science Institute,
which is operated by the Association of Universities for Research in
Astronomy, Inc., for NASA, under contract NAS 5-26555. PC gratefully
acknowledges the support of the Miller Institute for Basic Research,
and thanks the Institute for Advanced Study for their hospitality
where a portion of this work was conducted.






\end{document}